\documentclass[doublecol]{epl2}
\usepackage{hyperref}
\usepackage{amsmath} 
\usepackage{amsfonts} 
\usepackage{amssymb}
\usepackage{autonum}

\def\D{{\bf D}} 
\def\E{{\bf E}}  
\def\div{{\,\rm div}}
\def\DD{d_\ell}
\def\n0{\bar \lambda}
\def\Dl{{\bf D}_{\ell}}

\def\c0{{\lambda}}
\def\lagrange{\tau}

\title{A fluctuation-corrected functional of convex Poisson-Boltzmann theory} \shorttitle{Fluctuation corrections
in Poisson-Boltzmann theory} %Insert here a short version of the title if it exceeds 70 characters

\author{R. Blossey\inst{1} \and A.C. Maggs\inst{2}} \shortauthor{R. Blossey \etal}

\institute{\inst{1} University of Lille, Unit\'e de Glycobiologie Structurale et
  Fonctionnelle, CNRS UMR8576, 59000 Lille, France\\
  \inst{2} CNRS UMR7083, ESPCI Paris, PSL Research University, 10 rue Vauquelin, Paris, 75005,
  France }
  

\abstract{Poisson-Boltzmann theory allows one to study soft matter and
  biophysical systems involving point-like charges of low valencies. The inclusion 
  of fluctuation corrections
    beyond the mean-field approach typically requires the application of loop
  expansions around a mean-field solution for the electrostatic potential
  \(\phi({\bf r})\), or sophisticated variational approaches. Recently,
  Poisson-Boltzmann theory has been recast, via a Legendre transform, as a
  mean-field theory involving the dielectric displacement field
  \({\bf D}({\bf r})\). In this paper we consider the path integral formulation of
  the dual theory. Exploiting the transformation between \(\phi\) and \({\bf D}\),
  we formulate a dual  Sine-Gordon field theory in terms of the displacement
  field and provide a strategy for precise numerical computations of free energies  
  beyond the leading order.}
        
\begin{document}

\maketitle

\section{Introduction}

The description of electrolytes in soft-matter is commonly based on the
Poisson-Boltzmann equation, a mean-field theory for the electrostatic potential
\(\phi({\bf r})\) (see, e.g., \cite{podgornik}, and references therein). Formally, this theory is derived from
the partition function of a Coulomb system by performing a Hubbard-Stratonovich
transformation to an (imaginary) fluctuating field \(\phi({\bf r})\), later
identified with the electrostatic potential~\cite{netz99,netz01}.  This approach
is valid in the weak fluctuation regime, in which the saddle-point of the action
yields the standard Poisson-Boltzmann equation, which can be systematically
improved by a loop expansion~\cite{netz01,naji13}. An alternative approach to go
beyond mean-field theory is to invoke a variational
approach~\cite{netz03,buyukdagli16}. Going beyond mean field in Poisson-Boltzmann is
known to improve many features of the theory, for instance dielectric contrast
and image-charge interactions are much better described in theories which
allow fluctuations~\cite{xu}.

A difficulty with these approaches is that the action of the functional integral is
complex. A Wick rotation gives a real theory, but at the price of rendering the
effective free energy concave.  Thus, when the electrostatics is coupled to
other degrees of freedom, the extremization of the free energy becomes a
numerically difficult operation. Extrema result from a combination of minima in
the non-electrostatic degrees of freedom and maxima in the electrostatic
potential. This complicates many numerical studies of biophysical molecules in
which the solvent is described by Poisson-Boltzmann theory, and also has
repercussions for the validity limits of theories including fluctuation effects
beyond Poisson-Boltzmann theory~\cite{derek}.  This technical problem has been
circumvented~\cite{minimizing,dynamics} by a reformulation of the Poisson-Boltzmann
theory in terms of purely convex functionals, which is achieved by the means of
a Legendre transform, with the complex field \(\phi({\bf r})\) being
systematically replaced by the dielectric displacement field \({\bf D}({\bf
  r})\). Expressed in terms of the field \({\bf D} ( {\bf r } ) \), the resulting
theory yields a convex functional so that standard minimization techniques can
be applied, and loop corrections or variational approaches be defined as for the
original Poisson-Boltzmann theory. Though it was constructed only from the
mean-field, the one-loop correction in the dual theory (very surprisingly) has
been shown to yield the same fluctuation spectrum as obtained within the usual
formulation of Poisson-Boltzmann theory~\cite{stiffness}.

This approach, however, working from conventional mean-field theory, does not
help in constructing a systematic improvement in the dual formulation of the
theory. The aim of the present article is to formulate a systematic method for going beyond 
the ``one-shot'' approximation that has been considered until now for constructing
dual-theories and present a general formulation that is equivalent to the exact,
discretized statistical mechanics of the electrolytes that are formulated as
integrals over an electrostatic potential. We stress the generality of our approach, 
even if in this paper we chose to apply the formulation to only the simplest two-component
electrolyte.

With this motivation in mind, we revisit the dual approach from a functional integral
perspective which relies on the introduction of the dual transform via delta functions.
The arising transformation integrals can be performed either in an exact way, or
in a systematic fashion by invoking the Poisson summation formula. The latter 
will first give us access to a saddle-point like limit, but it equally allows us - with
no more than simple integration and one-dimensional Fourier transforms - to
explicitly calculate the corrections beyond the saddle point which arise from
the discrete nature of the charges. In this way we obtain a novel formulation of
Poisson-Boltzmann theory in terms of its dual which is exact. Our final result is a discretized 
free energy function, which, when sampled gives the \textit{exact} discretized energy of a set 
of discrete charges interacting with the Coulomb interaction. 

We expect that this novel dual formulation will be highly useful for the finite temperature {\it
simulation\/} of coupled conformational/electrostatic degrees of freedom,
rather than the minimization envisaged with the original mean-field theory. For this
purpose we formulate the theory in a formulation adequate for numerical computations on
lattices.   

The paper is organized as follows. In the second section, we discuss the notion
of duality, recall the formulation of the duality transform for the Poisson-Boltzmann 
functional and the derivation of the dual mean-field functional. In section three we 
derive a prescription to formulate the functional integral of a dual theory including
fluctuations beyond mean field.

\section{Duality in Poisson-Boltzmann theory}

We first recall the notion of duality previously employed in the context of the
Poisson-Boltzmann theory at the mean-field level~\cite{minimizing}.  We start
with the usual expression~\cite{netz99} for the partition function for a set of charged
particles interacting through the Coulomb interaction, described through a
fluctuating potential field \(\phi\):
\begin{equation}
  Z = \int D[\phi] e^{- \beta \int d^3{\bf r} h(\phi)}
\end{equation}
with
\begin{equation}
  h(\phi) = \varepsilon \frac{{(\nabla \phi)}^2}{2} + g(i\phi) - i \varrho_f \phi\, ,
\end{equation}
where \(g(i\phi)\) is the grand potential, while \(\varrho_f\) is the density of
fixed charges, which are typically confined to a small part of the system.
Introducing the electric field \({\bf E}\) with the help of a delta-function (and
being free with non-essential normalisation factors)
\begin{equation}
  Z = \int D[\phi] D[{\bf E}]e^{- \beta \int d^3{\bf r} h(\phi)} \delta({\bf E} + \nabla \phi) 
\end{equation}
yields after Fourier representation of the delta function with multiplier \({\bf D}\)
\begin{equation}
  Z = \int D[\phi] D[{\bf E}] D[{\bf D}] e^{- \beta \int d^3{\bf r} h(\phi, {\bf E}, {\bf D})} 
\end{equation}
with
\begin{eqnarray}
  h(\phi, {\bf E}, {\bf D}) & = & \varepsilon \frac{{\bf E}^2}{2} + g(i\phi) - i{\bf D} \cdot ({\nabla \phi}+ {\bf E}) - i \varrho_f \phi \\ 
                            &=& \varepsilon \frac{{\bf E}^2}{2} + g(i\phi) - i{\bf D}\cdot{\bf E}  + i\phi (\mbox{div} {\bf D} - \varrho_f) \nonumber\, .
\end{eqnarray}
Careful consideration of boundary conditions shows that
\begin{equation}
  \oint \phi \D.d{\bf S} = 0
\end{equation}
so that one requires either \(\phi=0\) or the normal component of the field
\(\D_n=0\) at the boundaries.  Performing the integration over the electric field
\({\bf E}\), one finds
\begin{equation} \label{pI}
  Z = \int D[\phi] D[{\bf D}] e^{- \beta \int d^3{\bf r} h(\phi, {\bf D})} 
\end{equation}
with
\begin{equation}
h(\phi,{\bf D}) = \frac{{\bf D}^2}{2\varepsilon} + g(i\phi) +i\phi(\mbox{div} {\bf D} - \varrho_f)\,.
\end{equation}
In this expression, \(\mbox{div}{\bf D} - \varrho_f \equiv s\) can be read as a Fourier transform
variable such that we arrive at the formal expression
\begin{equation} \label{intD}
h(\phi,{\bf D}) = \frac{{\bf D}^2}{2\varepsilon} - \ln\{{\cal F}(e^{g(i\phi)})\}[\mbox{div} {\bf D} - \varrho_f]\,.
\end{equation}
with \({\cal F}\) the Fourier operator. We emphasise that this step requires the
evaluation of just a \textit{single one-dimensional Fourier transform}. In numerical
applications this transform can be done once and tabulated, even if the
model for \(g(i \phi)\) is relatively complicated analytically. As emphasized
in~\cite{rudi} the function \(g(\phi)\) is typically a grand potential (or
pressure) of the \textit{uncharged} fluid.

We then find the main result of this paper
\begin{equation}
Z = \int D[{\bf D}] e^{- \beta \int d^3{\bf r}\, h_{eff} ({\bf D})}\,.
\end{equation}
An exact sampling of this partition function can then be performed either by
Monte Carlo simulation \cite{hybrid}.

In our formulation of the full statistical mechanics of the dual theory it is a
crucial feature that the duality transformation \(\phi \rightarrow {\bf D}\)  has
been carried out solely via the introduction of delta functions. Only this
approach guarantees the identity of the fluctuation spectra of the dual theory
as discussed in detail~\cite{fradkin85} in a general field theory setting. By
contrast, theories derived from reparametrizations are called pseudo-dual
theories~\cite{fradkin85} for which this property in general does not hold.
This is e.g.\ the case for the reparametrizations of the Poisson-Boltzmann theory
recently developed in~\cite{jadhao13, solis13} with the aim of defining convex
functionals for the electrostatic potential. These theories fail to produce the
correct fluctuation spectra as was shown explicitly in the calculation of the
one-loop correction~\cite{stiffness}.  Even in lattice gauge theories, dual
formulations can have similar technical advantages of positivity and convexity
and are an active research topic~\cite{fieldtheory}.

The above presentation has been at a formal level. In order to be explicit in this program we will 
now present a detailed calculation of eq.(\ref{intD}) for the case of a simple model for the grand potential, 
\begin{equation} 
  g(i\phi) = - 2 \c0 \cos(\phi\, {{e \beta}} ) \label{eq:cos}. 
\end{equation}
corresponding to a perfect gas model of the ions.  The full partition function
corresponding to~(\ref{eq:cos}) (without the complex coupling to the external
charge \(\varrho_f\)) is the well-known Sine-Gordon energy,
In three dimensions it has been widely studied due to its links to electrodynamics. 
Unlike the closely related model of charged hard-spheres there is no phase
transition \cite{kosterlitz}. In its discretized form the theory is regularized
by the lattice spacing.

For both the purpose of regularizing and simulating this theory, we will first discretize it.
Our dual model then is, by construction, entirely equivalent to this full, discrete Sine-Gordon 
system, as it includes all the fluctuations beyond mean-field theory that are contained in this starting theory.

% The mean-field functional is then given by
% \begin{equation} \label{dual1} {\cal F}_{dual}[{\bf D}] = \int d{\bf r}
%   \left(\frac{{\bf D}^2}{2\varepsilon} + {k_B}T\tilde{g}(\mbox{div} {\bf D} -
%     \varrho_f)/ec_0\right)
% \end{equation}
% where the Legendre transform of the hyperbolic cosine has the functional form
% \begin{equation} \label{dual2}
% \tilde{g}(s) = s \sinh^{-1}(s/2) - \sqrt{4 + s^2}\,.
% \end{equation}

\section{Discretization of the dual theory}

We now turn to a definition and calculation of the functional integral (\ref{pI}), for
the explicit case of the symmetric electrolyte given by (\ref{eq:cos}). 
The continuum action for the potential that we start with is
\begin{equation}
  h(\phi) = \varepsilon \frac{{(\nabla \phi)}^2}{2}  - 2 \c0 \cos(\phi\,
  {{e \beta}} )  - i   \varrho_f \phi\,, \label{eq:full}
\end{equation}
We discretize this model at a spacing  $\ell$ to a simple cubic lattice. The
potential and charges are associated with the sites of the lattice. Derivatives
are replaced by lattice differences, \( \DD \). The discretized energy corresponding to 
eq.~(\ref{eq:full}) is then
\begin{equation}
  h_{\ell}(\phi) = \ell \varepsilon \frac{{(\DD \phi)}^2}{2}  - 2\c0 \ell^3 \cos(\phi\,
  {{e \beta}} )  - i \ell^3 \varrho_f \phi\,. \label{eq:full2}
\end{equation}
We now change the integration variable from  \(\phi\) to the scaled
potential \(\phi/(e\beta) \). Giving the (a-dimensional) statistical weight
\begin{equation}
 \beta h_{\ell}(\phi) = \sum_{\text{sites}} \left [  \frac{\ell \varepsilon}{e^2 \beta} \frac{{(\DD
     \phi)}^2}{2}  - 2 \n0  \cos(\phi)  - i \bar \varrho \phi\right ] \,, \label{eq:full3}
\end{equation}
where we define \(\n0 = \beta \ell^3 \c0 \) and \(\bar \rho =  \varrho_f \ell^3/e \). We
note that \( {\ell \varepsilon}/{e^2 \beta} = \ell/(4 \pi \ell_B )\), with
$\ell_B$ the Bjerrum length.  To define the dual formulation we now consider the
discrete electric field, $\E_\ell = - \DD \phi$, to be associated the links of
the lattice and impose this constraint using a delta-function integral over
\(\Dl \), which is also considered as  link variable. The entire formulation
follows very closely the continuum calculation given above. We need only to
define the discrete equivalent of the divergence operator from the adjoint (or
matrix transpose) of the difference operator \(\DD\):
\( \div \rightarrow -\DD^T \)~\cite{yee}. 

% \begin{figure}[htb]
% \begin{center}
%   \includegraphics[height=5cm]{Figure/figure1.pdf}
% \end{center} 
%   \caption{Three length scales are important for the description of
%     the discretized electrolyte. The smallest $a=c_0^{-1/3}$ is the separation
%     between ions within the solution; $c_0$ is the ion concentration
%     of each species. We construct a discretization
%     such that on average there are $\n0>1$ ions within a cell. The
%     Debye length $\ell_D $ then forms a third, even larger scale over which
%     density correlations  are screened so that $\ell_D >\ell>c_0^{-1/3}$.\label{fig:figure1}}
% \end{figure}

The partition function in its discretized form is thus
\begin{equation} \label{pI2}
  Z = \int D[{\Dl}] D[\phi]\, e^{- \sum
\left(\frac{{e^2 \beta  }}{2 \ell \varepsilon} {\Dl}^2 + i(-\DD^T {\Dl} - \bar \varrho )\phi - 2\n0 \cos (\phi) \right)}\, .
\end{equation}
The expression can be reorganized into
\begin{eqnarray}
  Z & = & \int D[{\Dl}]\, e^{-\sum \frac{{e^2 \beta
          }}{2\ell \varepsilon} {\Dl}^2 } \int D[\phi] e^{\sum\left(-is \phi + 2\n0\cos(\phi)\right)} \nonumber \\
    & = & I[{\Dl},J[\phi]]  \label{phi-integral}
\end{eqnarray}
with $s \equiv -\DD^T {\Dl} - \bar \varrho$. Here sums in the exponentials are
understood as being over sites for potentials and charges and over links for
field components. It is here that we see explicitly that the only step that is
needed to complete the program is the integration over the field \(\phi\).
%\begin{equation} \label{phi-integral}
%\int D[\phi] e^{\sum\left(-is \phi + 2\n0\cos(\phi)\right)}\,.
%\end{equation}
As the field $\phi$ in the expression (\ref{phi-integral}) is purely local, we
can drop the functional notation and thus have to deal with a simple integral
for each site of the discretization
\begin{equation} \label{eq:z1} 
z(s) =  \int d\phi\, e^{\left(-is \phi + 2\n0\cos(\phi)\right)} \,. 
\end{equation}
For a general model of electrolyte we can expect that this integral will be difficult to perform analytically, but it
can always be treated in a numerical manner, in order to generate the
\textit{exact} effective theory for $\Dl$. We can make further analytic progress
for the case of the two-component electrolyte, which we now present in detail.

\section{Two-component electrolyte}

The standard way of proceeding to calculate $z(s)$ would be to invoke complex integration.
The argument of the exponential function has zeroes at $2\n0 \sin(\phi) = is $,
hence at $\phi = \sin^{-1}(is/2\n0)$, whereby the inversion of the sine function
fixes the saddle to be retained. Here, we will work through the saddle-point evaluation
and its corrections in a different manner, in view of our interest to have an approach which is adapted
to numerical computations. 

The mathematical tool we use in our formalism is the Poisson summation
formula which expresses a sum over integer occupation numbers by an equivalent
sum over Fourier coefficients. This gives a mathematically rigorous formulation
in which the dominant contribution is the mean-field free energy already but
gives a framework in which higher-order corrections to the dual theory are
expressed as an exact Fourier series, with exponential convergence.  Thus we
now find an analytic expansion for eq.~(\ref{eq:z1}). 

In order to implement this program we note that we first rewrite \(z(s) \) in the discrete form 
\begin{equation} 
z(s) =  \sum_n \delta (s-n)\,g(n) =  \sum_n \delta (s-n)\,e^{-f_e(n)}  \label{eq:delta}
\end{equation}
where we then need to calculate the weighting function \(f_e(s)\). 
In order to perform this step we expand the exponentiated cosine in
eq.~(\ref{eq:z1}) as a Taylor series of the two complex exponentials using
\begin{equation}
  \exp{(\n0 e^{i \phi})  } = \sum_{n=0}^\infty \frac{e^{i n \phi      +n\mu}}{n!}\, , \label{eq:expand}
\end{equation}
where for convenience we define $\mu = \ln(\n0) $.  Substituting
eq.~(\ref{eq:expand}) in eq.~(\ref{eq:z1}) and using the definition of the delta function we find
\begin{equation}
z(s)=\sum_{n_1,n_2=0}^{\infty}\frac{1}{n_1!} \frac{1}{n_2!} \, e^{\mu (n_1 +n_2)} \,\delta ( s + n_1 -
  n_2 ) \label{eq:partsum}
\end{equation}
where $n_1$ and $n_2$ represent the occupation numbers of positive and negative ions.  

In order to perform the summation, we can invoke the Poisson summation formula which
tells us that if $g(n,s)$ is a function defined on integers and $\tilde g$ is its continuous Fourier 
transform
\begin{equation}
 z(s)=\sum_n g(n,s) = \sum_k \tilde g(2 \pi k,s) =\sum_k z_k(s)\,  \label{eq:poisson}
\end{equation}
where $z_k(s)$ is the contribution to the $k$-th Fourier mode to the total sum $z(s)$.
In order to apply this identity we solve the delta-function constraint and reduce the problem to a 
single summation, which leads to the expression
\begin{equation}
z(s) = \sum_n g(n,s) = \sum_n \frac{1}{n!} \frac{1}{(n + |s|)!} e^{\mu ( 2n+|s| )}\,.     \label{eq:singlesum}
\end{equation}
This expression is a simple function of the variable $s$, which can be tabulated, or
interpolated to arbitrary precision for numerical work.

Some additional care is, however, needed for our problem: we here require the sum is over 
\textit{positive} integers, whereas the Poisson formula applies for a sum over \textit{positive and negative} 
integers. This can be achieved by imposing a multiplicative weighting function.

We implement this first for the $k=0$ contribution to the full partition sum eq.~(\ref{eq:poisson}). 
For convenience we go back to a continuous variable $n$ with which we can write this contribution as an integral 
over the summand of eq.~(\ref{eq:poisson}); it thus is the \textit{naive} integral over the occupation number variable, 
forgetting about the discrete nature of the elementary charges. It reads as
\begin{equation}
z_0(s) = \int_{-\infty}^\infty {\rm d}n\,  w(n)\, g(n,s)\,  
\end{equation}
where the weighting function is given by, e.g.,
\begin{align}
 w(n) = 
 \begin{cases}
    0\quad &\text{if} \quad  n < -1 \\
     \frac{e^{-1/n} } {e^{-1/n} + e^{1/(1+n)}} \quad &\text{if}  -1 < n < 0 \\
     1 &\text{if} \quad n > 0\,.
\end{cases}
\end{align}
This cross-over function is smooth and increases from zero to unity on the
interval \(-1<n<0\). The choice of $w(n)$ is non-unique; we note that the general theory
of Fourier analysis show that the final results \textit{does not depend on the
particular choice} we have made. We also note that the extension of the factorial 
function to the real numbers, which is needed to represent $g(n,s)$ in its continuous form,
is also non-unique. The commonly used extension is Euler's Gamma
function, but alternative extensions due to Hadamard are also
possible~\cite{hadamard}.  
 
The integral $z_0(s)$ is dominated for $\n0 > 1$ by a saddle.  We will also show that in
this limit the contributions \(z_k(s)\) for \(|k| \ge 1\) are exponentially
smaller and that to high accuracy \(z(s)\) can be replaced by \(z_0(s)\).
To study the nature of the saddle we work with an improved version of the Stirling formula
\begin{equation}
\ln(n!) \approx (n+1/2) \ln(n+1/2) - (n+1/2) +O(1) = S(n)\nonumber
\end{equation}
which correctly includes the term $(\ln{n})/2$ in its expansion. 
Replacing the single continuous variable $n$ again by 
two continuous variables $n_1$ and $n_2$ we then have the expression
\begin{equation}
  z_0(s) = \int_{n_1,n_2}\!\!\!\!\!\!\!\!\!\! \exp\left (\mu (n_1 + n_2) - S(n_1) - S(n_2) \right ) 
  \delta(s+n_1-n_2). \label{eq:saddle0}
\end{equation}
\\
The saddle-point is best studied by implementing the delta-function constraint with a 
Lagrange multiplier $\lagrange$, which leads to the following equations,
\begin{eqnarray}
  s =  &2 \n0 \sinh(\lagrange) \nonumber \\ 
  n_1+ 1/2= & \n0 e^{-\lagrange}\\
  n_2 +1/2= & \n0 e^{+\lagrange}\,. \nonumber
\end{eqnarray}
On eliminating $\lagrange$ we find the values of the occupation numbers
which extremize the integrand:
\begin{eqnarray}
  n_2 +1/2 = &\frac{ (\, s + {(s^2 + 4 \n0^2)}^{1/2})}2 \nonumber \\
  n_1 +1/2 = &\frac{(-s + {(s^2 + 4 \n0^2)}^{1/2})}2\,.
\end{eqnarray}
When  \((s/\n0)\) is small both $n_1$ and $n_2$ equal the background
occupation number $\n0$. The value of the saddle-point eq.~(\ref{eq:saddle0}) is:
\begin{align}
  \ln(z_0) \approx 
  s \ln \left [\frac{s}{2\n0} + \sqrt{1+{(s/2\n0)}^2} \right ] \nonumber
  \\
  -\sqrt{s^2+4 \n0^2} - \ln{(\n0)} \label{eq:saddle}
\end{align}
and we define \(f_0(s)= -\ln{z_0}\). With $s$  small \(z_0 =O(e^{2\n0})\).

We now evaluate the second derivative, \(\Delta(s)\),  of the action at the saddle 
\begin{equation}
  \Delta(s)= \frac{1}{n_1+1/2} + \frac{1}{n_2+1/2}  =  \frac{1}{\n0^2} \sqrt{4 \n0^2 + s^2}\,.
\end{equation}
For small $s$, $\Delta(s) = 2/\n0$.  At the saddle-point we have the
quadratic approximation to the statistical weight,
\begin{equation}
  f(n,s)= f_0(s) + \frac{1}{2}  \Delta\,  {(n-n_s)}^2, \quad n_s = \mbox{min} (n_1,n_2)\,.			\label{eq:quad}
\end{equation}
According to the sign of $s$ we use the smaller of $n_1$ and $n_2$ as the
primary variable mean-field occupation number.
Performing the integral over the full occupation number $n$ in
eq.~(\ref{eq:quad}) we find that the effective action for eq.~(\ref{eq:delta}) is
\begin{equation}
  f_e(s)= f_0(s) + \frac{1}{2} \ln(\Delta(s))\,. 			\label{eq:loop}
\end{equation}
%\AM{OK, if \(s\) integer, the result should be I think
%\begin{equation}
% f_e(s)=\sum_p \delta (s-p) \left [ f_0(p) + \frac{1}{2} \ln(\Delta(p)) \right]\,. 	
%\end{equation}
%}

%\AM{The technical point that remains is transforming this into an integral
 % rather than a sum when inserted into the final partition function as an
 % integral over D}

\begin{figure}[htb]
  \includegraphics[height=12.5cm]{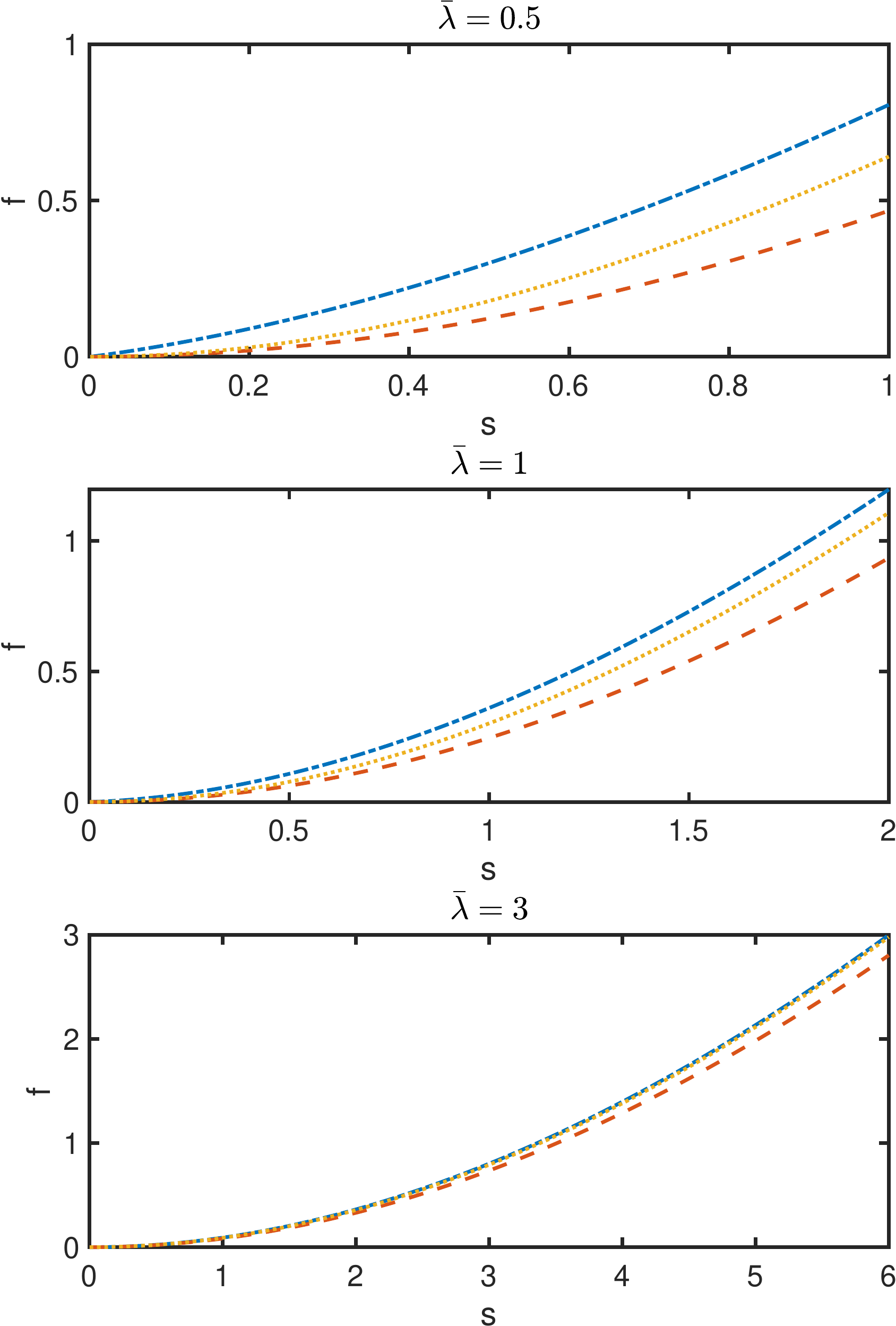}
  \caption{Comparison of the evaluation of the sum eq.~(\ref{eq:partsum}), (blue dot-dash) compared
    to the saddle point approximation eq.~(\ref{eq:saddle}), (red dash) and the quadratic
    approximation, eq.~(\ref{eq:loop}), (yellow dotted). Three different values of $\n0$. For
    already modest values of $\n0=3$ (bottom) the quadratic approximation approximation is already excellent with a good overlay
    of the exact and approximate curves. For $\n0<1$ (top) the mean field and quadratic approximations
    underestimate the true free energy. Curves shifted so that $f(0)=0$.
    \label{fig:saddle}}
\end{figure}

With the help of the Poisson summation formula we now calculate the \textit{higher Fourier contributions to the sum} 
eq.~(\ref{eq:poisson}) which come from the discrete nature of the elementary charges. For
$\n0>1$ the Fourier components are also calculated in the saddle-point
approximation. We take the quadratic approximation to the energy
eq.~(\ref{eq:quad}) and regroup the positive and negative Fourier coefficients:
\begin{equation}
  \tilde z_k= (z_{k} +z_{-k}) = e^{-f_0(s)} \int 2 \cos{(2 \pi  k n)} e^{-\Delta{(n-n_s)}^2/2} {\rm d}n
\end{equation}
which results in the expression
\begin{equation}
  \tilde z_k = \frac{e^{-f_0(s)}}{\sqrt{\Delta}} 2\cos(2 \pi k n_s) e^{-2 k^2
    \pi^2/\Delta}\,.
\end{equation}
If we consider that $\Delta = O(2/\n0)$ then the amplitude of this contribution is
\begin{equation}
  \tilde z_k/z_0 = O(  e^{-\pi^2k^2 \n0})\, .
\end{equation}
If $\n0 \gtrsim 1$ then this contribution is strongly suppressed in comparison
with the $k=0$ contributions to the partition function.

We now calculate the importance of the crossover function $w(n)$ which
constrains the sums and integrals to be over positive occupations. Near $n=0$
the function $\ln(n!) = O(1)$. The Fourier transform of the weighting function
can be estimated by a further saddle-point calculation to be
$O(1)\times e^{-\sqrt{k}} $,~\cite{johnson}. We thus expect that the
contribution to the partition sum coming from the constraint of the sum over
positive occupation numbers is of $O(1)$.

Finally, by a regrouping of the contributions we conclude that an expansion for
the Fourier partition sum eq.~(\ref{eq:poisson}) is given by the formula 
\begin{equation}
z(s) = \frac{e^{-f_0(s)} }{\sqrt{\Delta(s)}} (1+2 \cos(2 n_s) e^{-2 \pi^2/\Delta(s)}) +O(1)
\end{equation}
where $-f_0(s) \sim 2 \n0 $ and we have kept just the first and largest
non-trivial Fourier component for $\tilde z_k$.\footnote{We have also assumed
that the background charge density is not so large that it overwhelms
the electrolyte concentration.}

We conclude that when the discretization is such that $\n0 >1$, we can drop the
oscillating terms as well as the end-point corrections to find the simplified 
expression
\begin{equation}
  f_e (s) = f_0(s) + \frac{1}{4} \ln\left(\frac{s^2+4 \n0^2}{\n0^4}\right)\, . 
  \label{eq:saddle2}
\end{equation}
This result is to be compared to the exact sum eq.~(\ref{eq:partsum}) in
Fig.~\ref{fig:saddle}. We see that the the analytic expression is an excellent
fit to the full numerical evaluation.

Thus for $\n0>1$, the final \textit{effective functional for the displacement
  field} $\Dl$ is given by the
\begin{equation}
  h_{eff} = \sum_{\text{links}} \Big [ \frac{e^2 \beta}{\ell \varepsilon}
  \frac{\Dl^2}{2} + f_e(- \DD^T \Dl - \bar \varrho) \Big ] \label{eq:Aeff}
\end{equation}
where \(f_e(s)\) is given by eq.~(\ref{eq:saddle}) and eq.~(\ref{eq:saddle2}),
{and the measure includes the delta-functions of eq.~(\ref{eq:delta})}. 
This functional is
valid beyond mean-field as it includes the fluctuations in the underlying field
$\phi$, it is also explicitly an energy function that should be sampled by Monte
Carlo or molecular dynamics.

% \subsection{Low density expansion}
% When \(\n0\) is small only a small number of terms in the sum eq.~(\ref{eq:partsum}) contribute. We can
% find an  approximation to \(z(s)\) by taking just the very first term in the series:
% \begin{equation}
% z= \frac{e^{\mu s }}{\Gamma(1+s)  }; \quad \quad f=-\log{z} \label{eq:low}
% \end{equation}
% The result of this approximation is shown in Figure~\ref{fig:truncation}. It can
% be seen that the result works very well. However, this limit does not have
% equivalent of the Poisson formula, where we replace summation over the partition
% function by an integral. We have also checked that at intermediate values of
% $\n0$, the function $z(s)$ can be replaced by a simple interpolation to very
% high accuracy.

% \begin{figure}[htb]
%   \includegraphics[height=12.5cm]{Figure/figure3}
%   \caption{Comparison of the evaluation of the sum eq.~(\ref{eq:partsum}), (blue
%     dot-dash) compared to the approximation of taking the very first term in the
%     summation, eq.~(\ref{eq:low}) (red dots). The approximation of keeping two
%     terms in the series is shown in yellow dash.  Curves shifted so that
%     $f(0)=0$. Three different values of $\n0$. The approximation of a single
%     term is already excellent for $\n0 \sim 0.2$. }\label{fig:truncation}
% \end{figure}

\section{Conclusions}
 
In this Letter we have given an explicit formulation of the duality-transformed
Poisson-Boltzmann theory, for the illustrative case of a system of a symmetric
electrolyte. The partition function is given by a double-integral over the
potential field variable $\phi$ and the dielectric displacement field
${\Dl}$. The potential integral can be evaluated as a systematic series with the
help of the Poisson summation formula. The result is a Fourier series for the
\(\phi\) integral. for \(\n0>1\) we evaluated this series analytically, but the
Fourier transforms can also be evaluated numerically for small \(\n0\). The
final result is a theory which has physical content identical to the Sine-Gordon
model of a charged lattice gas, which goes beyond mean field theory and
\textit{includes all fluctuations in an exact manner}. The final theory remains
clearly analytically intractable. However, it is of now of a form which is
simple to simulate, and to couple to external charges. This is in contrast to
the original field-theory formulation in terms of the potential, which included
a complex coupling to external charges. This opens the possibility of simulation
of complex coupled situations where electrostatic fluctuations are important.

Although we have derived our formulation for the case of the symmetric
electrolyte our approach is not limited to this choice and should allow the
treatment of more complex models of electrolyte, fully including fluctuations
beyond mean-field theory in their coupling to external soft-matter systems.

\bibliographystyle{eplbib}
\bibliography{conv}

\end{document}